\documentclass[12pt]{article}
\usepackage{graphics,graphicx,epsfig}
\usepackage{amssymb,amsmath,amsfonts}
\usepackage{ifthen}

\def\hbar{\hspace{0pt}\raisebox{1pt}{$-$} \hspace{-7pt} h}

\def\5{\overline 5}

\newcommand{\be}{\begin{equation}}
\newcommand{\ee}{\end{equation}}
\newcommand{\bea}{\begin{eqnarray}}
\newcommand{\eea}{\end{eqnarray}}

\newcommand{\lsim}{\,\raise.3ex\hbox{$<$\kern-.75em\lower1ex\hbox{$\sim$}}\,}
\newcommand{\gsim}{\,\raise.3ex\hbox{$>$\kern-.75em\lower1ex\hbox{$\sim$}}\,}
\newcommand{\LL}{\mathcal{L}}

\newcommand{\TeV}{\text{ TeV }}
\newcommand{\GeV}{\text{ GeV}}

\newcommand{\bino}{\tilde{B}}
\newcommand{\wino}{\tilde{W}}

\newcommand{\higgsino}{\tilde{H}}

\begin{document}

\begin{titlepage}
\renewcommand{\thefootnote}{\fnsymbol{footnote}}
\setcounter{footnote}{0}
\begin{flushright}
SLAC--PUB--10489\\
June 2004\\
hep-ph/0406144\\
\end{flushright}
\vskip 1cm
\begin{center}
{\large\bf Dark Matter in the Finely Tuned Minimal Supersymmetric 
Standard Model}
\vskip 1cm
{\normalsize
Aaron Pierce$^{1,2}$\footnote{AP is supported by the U.S. Department of Energy 
under contract number DE-AC03-76SF00515.}\\
\vskip 0.5cm
1. Theory Group, Stanford Linear Accelerator Center\\
   Menlo Park, CA 94025\\
2. Institute for Theoretical Physics, Stanford University\\
   Stanford, CA 94305
\vskip .1in
}
\end{center}
\vskip .5cm

\begin{abstract}
We explore dark matter in the Finely Tuned Minimal Supersymmetric
Standard model recently proposed by Arkani-Hamed and Dimopoulos.  
Relative to the MSSM, there are fewer particles at 
freeze-out, so the calculation of the relic abundance simplifies.  
Similarly, the predictions for direct detection of the dark matter 
sharpen.  There is a large region of mixed bino--higgsino
dark matter where the lightest supersymmetric particle will be 
accessible at both the LHC and future direct detection experiments,
allowing for a conclusive identification of the dark matter particle.  
Typical dark matter-nucleon cross sections are 
$10^{-45}-10^{-44}$ cm$^{2}$. This model also possesses a novel region 
where the dark matter annihilates via an $s$-channel Higgs boson 
resonance.
\end{abstract}

\vskip .5cm
\end{titlepage}

%
\section{Introduction}
\renewcommand{\thefootnote}{\arabic{footnote}}
\setcounter{footnote}{0}

In the Minimal Supersymmetric Standard Model (MSSM)\cite{MSSM}, new 
particles are placed at the weak scale to stabilize the 
Higgs boson mass hierarchy.  This theory has two important successes: 
gauge coupling unification \cite{GCU} and the 
presence of a natural dark matter candidate\cite{MSSM,Goldberg}.  Both  
features arise from new particles at the weak scale.  
Once R-parity is imposed, the existence of dark matter and the unification 
of gauge couplings can be viewed as predictions of the theory.  

Recently, Arkani-Hamed and Dimopoulos argued that in light of the 
cosmological constant problem, issues of naturalness should be 
treated delicately.  The end result of this argument was a
finely-tuned minimal supersymmetric standard
model where the higgsinos and gauginos are kept light 
by chiral symmetries, while all scalars are ultra-heavy\cite{NimaSavas}.  
While this model solves many 
of the problems associated with the MSSM, from the low-energy perspective 
the lightness of the standard model-like Higgs is accomplished via a 
fine-tuning.  This scenario was recently christened ``split supersymmetry'' 
by Giudice and Romanino.  

Since this model is unconcerned with softening the divergence in 
the Higgs boson mass, it seems at first unclear what masses 
to expect for the new fermions.  Luckily, a hint is given by 
the existence of dark matter.  A stable, 
weakly coupled particle gives a cosmologically interesting 
abundance only if its mass is at the weak scale\cite{LeeWeinberg}.  
After the gauginos have been placed at the
weak scale to give the dark matter, gauge coupling unification becomes
as a successful ``prediction'' of the theory.  In split supersymmetry, 
dark matter provides the sole link between the masses of the new 
particles and the weak scale.  Thus,  understanding the dark matter has 
strong implications for both future collider searches and for direct
dark matter detection experiments.  We find that there is a large region
of parameter space accessible to both future direct detection experiments 
and the Large Hadron Collider (LHC).

\section{Relic Abundance}
At first, it might seem possible to compute the dark matter relic density
in split supersymmetry by taking a MSSM relic abundance 
calculation\cite{JKG}, and simply eliminating all diagrams containing the 
(now decoupled) scalars.  
However, the situation is more subtle: there are several differences 
between the low-energy physics of split supersymmetry and the MSSM with the 
scalars eliminated.  These should be accounted for in an accurate 
relic abundance calculation.  

Perhaps the most salient differences between the MSSM and split 
supersymmetry are in the properties of the lightest Higgs boson.  
In the MSSM, the quartic coupling of the Higgs boson is related to the 
standard model gauge couplings: $\lambda = (g^{2} + g'^{2})/8$, leading to the 
tree level relation  
$m_{h}^{2} =M_{Z}^{2} \cos^{2} 2 \beta$.  In split supersymmetry, this 
relation exists only above the masses of the scalars, $m_{S}$.  
Below this scale, the quartic 
coupling flows
away from its supersymmetric value.  The result is that the 
Higgs boson may be as heavy as 170 GeV for large values 
of $m_{S}$ \cite{NimaSavas,Stanford}.  Because of the large mass, the 
width of the Higgs will be substantially different.  At 130 GeV, a standard 
model-like Higgs has a width of a few MeV; at 170 GeV, the width 
approaches 1 GeV.  Even for $m_{h} > 130$ GeV, the Higgs boson decays 
dominantly to $W W^{*}$, a situation not normally found in the MSSM.  
For our relic density and direct detection calculations, we 
utilize a modified version of 
DarkSUSY\cite{DarkSUSY}.  DarkSUSY 
only incorporates two-body final states; it was 
necessary to change the package to take into account the 
process $\chi^{0} \chi^{0} \rightarrow h \rightarrow W W^{*}$.  
 
Another difference between split supersymmetry and the MSSM is in 
the couplings between the gauginos, higgsinos, and the Higgs boson:
\begin{eqnarray}
\LL &\ni&  \bino (\kappa'_1 h^\dagger \higgsino_1 + \kappa_2' h \higgsino_2)
+\wino^a( \kappa_1 h^\dagger \tau^a \higgsino_1 + \kappa_2
\higgsino_2 \tau^a h).
\end{eqnarray}
In the MSSM, these couplings are related by supersymmetry to the 
gauge couplings:
\begin{eqnarray}
\label{eqn:kappas}
\nonumber &&\kappa'_1= \sqrt{\frac{3}{10}} g_1 \sin \beta
\hspace{0.3in} \kappa'_2= \sqrt{\frac{3}{10}} g_1 \cos\beta
 \\
\nonumber &&\kappa_1= \sqrt{2} g_2 \sin \beta \hspace{0.3in}
\kappa_2= \sqrt{2} g_2 \cos \beta.
\end{eqnarray}
In split supersymmetry, this relation only exists above the 
scale $m_{S}$.  Below $m_{S}$, 
these couplings must be run down to their low energy values, 
accomplished here by using the one-loop renormalization group 
equations of \cite{Stanford,Giudice}

The couplings of Eqn.~(\ref{eqn:kappas}) also feed into the 
off-diagonal entries of the chargino and neutralino mass matrices.  
They become:
\begin{eqnarray}
M_{{\chi}^{0}}=
\left( \begin{array}{cccc}
M_{1} & 0 & -\frac{\kappa_{2}^{\prime} v}{\sqrt{2}} & \frac{\kappa_{1}^{\prime} v}{\sqrt{2}}\\
0 & M_{2} & \frac{\kappa_{2} v}{\sqrt{8}} &
-\frac{\kappa_{1} v}{\sqrt{8}} \\
-\frac{\kappa_{2}^{\prime} v}{\sqrt{2}}
& \frac{\kappa_{2} v}{\sqrt{8}} &
0 & -\mu \\
\frac{\kappa_{1}^{\prime} v}{\sqrt{2}} & 
-\frac{\kappa_{1} v}{\sqrt{8}} & -\mu & 0
\end{array} \right), \hspace{.3in} 
M_{{\chi}^{\pm}}= \left( \begin{array}{cc}
M_{2} & \frac{v \kappa_{1}}{2} \\
\frac{v \kappa_{2}}{2} & \mu
\end{array} \right),
\label{massmat}
\end{eqnarray}
with $v= 246$ GeV.
Taking these modifications into account, it is possible to calculate the 
relic abundance of the Dark Matter.

\subsection{Bino Dark Matter}
The most interesting region for both colliders and direct detection involves 
a $\bino$-like LSP (Lightest Supersymmetric Particle), 
where $M_{1} < M_{2}, \mu$.   Assuming universal boundary 
conditions for $M_{1}$ and $M_{2}$ at the high
scale, we expect this condition to hold due to the renormalization
group evolution.

In the MSSM, there are many 
ways in which the $\bino$'s may annihilate in the early universe.  
For example, there exist regions in the MSSM parameter space where 
the $\bino$ can co-annihilate with a tau slepton or top squark.  
In split supersymmetry, however, the scalars are
quite heavy and are absent at the time of freeze-out due to the 
Boltzmann suppression.  Similarly, no ``funnel'' region exists where the 
LSP annihilates through the pseudo-scalar Higgs boson.  Also, there is 
no $t$-channel exchange of sleptons. In fact, in a theory where all 
scalars are decoupled, a pure bino is completely non-interacting.  Therefore, 
in split supersymmetry, the only relevant interactions of a $\bino$-like
LSP occur through the mixing of the $\bino$ with 
the $\higgsino$ and the $\wino$.

Fig.~1 shows the points with the correct relic abundance in the $\mu$-$M_{1}$ 
plane.  We have used the post-WMAP $2\sigma$ allowed region for the Dark 
Matter, $0.094 < \Omega_{DM}h^{2} < 0.129$ \cite{WMAP}.  We plot points 
satisfying the relic density constraint, starting with a base 
case of $\tan \beta=1$ (at the high scale), $m_{S}=10^{13} \GeV$, and imposing
the phenomenological relation $r \equiv M_{2}/M_{1}=2$ (at low 
energies).\footnote{It is natural to impose a
unified relation on the gaugino masses.  Where this condition is 
imposed is model dependent. One possibility is the grand unified scale. 
Another arises in the case of the Scherk-Schwarz breaking model 
of \cite{NimaSavas},  where the unified condition is imposed at the scale 
$1/R$.  
Given this model dependence, we choose to impose a 
phenomenological relation at the low scale.}
To get a feeling of how the
allowed region depends on these parameters, we show the allowed region 
when the scale of supersymmetry breaking is changed to 
$m_{S}=10^{6} \GeV$, when $\tan \beta$ is changed to 40 and when $r=4$.  
In the MSSM, $\tan \beta$=1 is not allowed due to the LEP limit on 
the Higgs boson mass, here there is no such difficulty. 
For much lower values, a 
Landau pole for the top Yukawa coupling would be encountered not far 
above the scale $m_{S}$.

\begin{figure}
\begin{center}
\epsfig{file=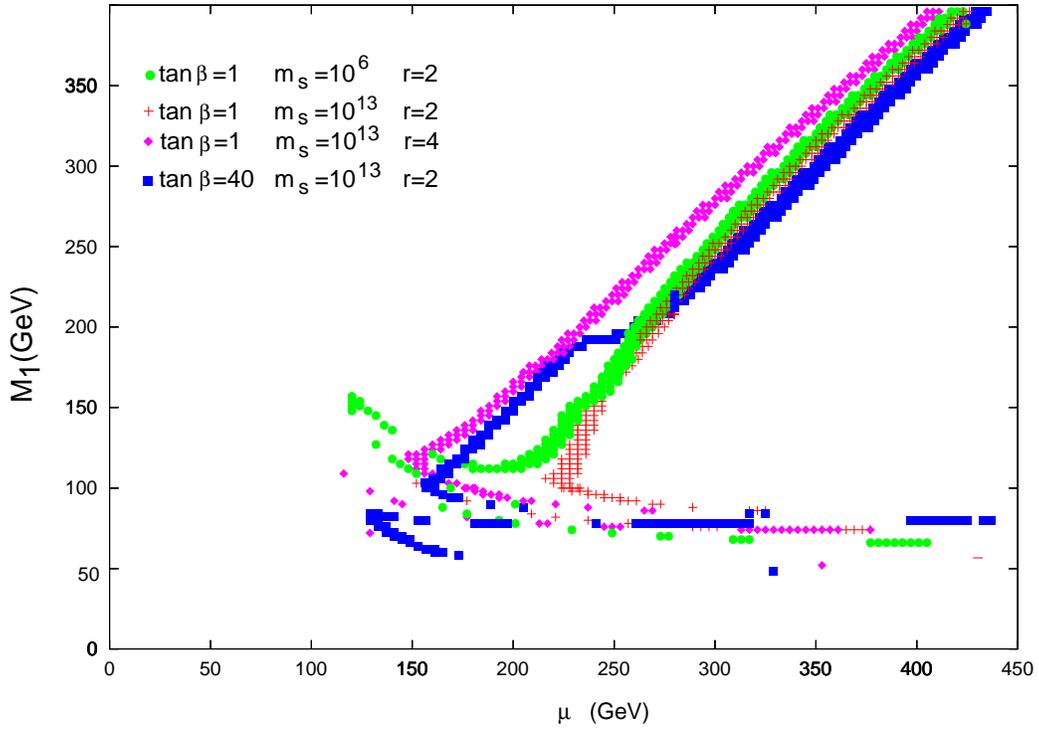,width=\columnwidth}
\label{fig:relic}
\caption{Points in the $\mu-M_{1}$ plane that satisfy the relic abundance
constraint from WMAP.  The region at low $M_{1}$ extending to large $\mu$ 
is the Higgs resonance region; the dark matter can be very nearly bino in this 
region. The diagonal line represents a mixed 
bino-higgsino dark matter region. The allowed region is shown for 
vaiour values of $\tan \beta$, scalars masses, $m_{S}$, and 
$r\equiv M_{2}/M_{1}$.  The top quark mass is set to 178 GeV.}  
\end{center}
\end{figure}

There are several distinct regions visible in the plot.  First, consider the 
diagonal stripe cutting across the parameter space. This is a region
of mixed bino-higgsino dark matter. Writing
\begin{equation}
\chi_{1}^{0}=N_{11} \bino + N_{12} \wino + N_{13} \higgsino_{u} + N_{14} \higgsino_{d},
\end{equation}
we can define the higgsino fraction 
$h_{f} \equiv |N_{13}|^{2} + |N_{14}|^{2}$.
Near the base of this stripe of parameter space,  
$h_{f} \sim .2$. 
In this region, the relic abundance is largely controlled by the 
process $\chi \chi \rightarrow W^{+} W^{-}$.  This process receives 
contributions from 
$t$-channel exchange of charginos, $s$-channel exchange of a $Z$ boson, 
and $s$-channel
exchange of the Higgs boson.  This region is nearly a 
straight line in the $\mu-M_{1}$ plane because 
all the above diagrams depend on $h_{f}$, which in turn can be written 
as a function of the slope $M_{1}/\mu$.  The top-quark 
threshold at $M_{1} \sim 175 \GeV$ is visible.  Note,
for $r=4$ the stripe is shifted somewhat, the larger $M_{2}$
affects the chargino masses, and thus the annihilation rate. 
This stripe of parameter 
space provides a region where the both the LSP and charginos can be 
visible at the LHC.  This stripe does connect continuously with an
experimentally difficult region of purely higgsino dark matter at 
high mass, which we will discuss momentarily.  The region 
above and to the left of this stripe the LSP does not provide all of the 
dark matter; another component would be needed to make up the remainder.

The horizontal bands at $M_{1} \sim 80$ GeV represent
a region where the LSP annihilates resonantly via a Higgs boson in
the $s$-channel. This region has special importance in 
split supersymmetry.  In the usual MSSM, the Higgs boson has a mass less 
than 130 GeV, and a tiny width, making resonant annihilation unlikely. 
In split supersymmetry, the Higgs boson can be much heavier, $\sim 160$ GeV, 
where the width approaches a GeV.  If the LSP has 
mass, $m_{\chi} \sim 0.5 m_{h}$, annihilation through the 
Higgs pole can be very efficient, and the composition 
of the dark matter can be very bino-like.  
In the resonance region, the relevant couplings for 
annihilation are $\kappa_{1}', \kappa_{2}'$, which must be run down 
from their supersymmetric values at $m_{S}$.  Looking at  
Fig.~1, the exact location of this region in the 
$M_{1}-\mu$ varies.  Comparing the region for 
$m_{S}=10^{13} \GeV$, $\tan \beta=1, r=2$ 
to $m_{S}=10^{13} \GeV$, $\tan \beta=40, r=2$ and 
$m_{S}=10^{13} \GeV$, $\tan \beta=1, r=2$ to  
$m_{S}=10^{6} \GeV$, $\tan \beta=1, r=2$ a dependence on 
both $m_{S}$ and $\tan \beta$ is visible.  This is because these 
parameters affect the Higgs mass.  In this region, the lightness of the 
LSP should allow it to be probed at the LHC. 

The area near the intersection of the diagonal stripe and the horizontal band
represents a particularly complicated region where many processes are at 
work contributing to the 
relic abundance.  Here, for our choice of $r\equiv M_{2}/M_{1} = 2$, it 
is possible to get Dark Matter that has a comparable fraction of 
wino, higgsino and bino.  If $r$ is increased, the wino fraction 
decreases.  Furthermore, co-annihilations can be in this region.  

Finally, for the $m_{S}=10^{13}, \tan \beta=40, r=2$ and   
$m_{S}=10^{13}, \tan \beta=40, r=4$ regions, a few points are visible 
at $M_{1}\sim 50 \GeV$.  There the $Z$-resonance becomes important.  These 
points are not visible for the lower values of $\tan \beta$ or $r$ because 
they are excluded by the chargino mass bound.  This region should
be easily accessible at colliders.

\subsection{Higgsino and Wino Dark Matter}
In the case of purely Higgsino dark matter, constraints 
from LEP force the LSP to be heavy enough so that the co-annihilation with 
charginos to $W^{\pm}$ bosons are allowed. So, a light Higgsino annihilates 
away too efficiently to make up the Dark Matter.  As the mass of the Higgsino
is increased, the $WW$ threshold is reached, and the problem worsens.  
Nevertheless, by making the Higgsino heavier, it is eventually possible to 
make up the observed dark matter, but this is always for $\mu > 1 \TeV$, the 
exact value is sensitive to $\tan \beta$.  In this case, it would be very 
challenging to observe the Dark Matter (or any signature of supersymmetry) 
at the LHC. 

A similar situation exists for wino dark matter.  It can (co)-annihilate
very efficiently to gauge boson(s), and only by making it heavy can the 
right relic abundance be achieved.  To achieve the correct relic abundance,
$M_{2} \sim 2.4$ TeV, making this a difficult scenario for colliders indeed.  
While this scenario is unlikely to occur in a model with unified boundary 
conditions, $r <1$ can  occur when the gaugino masses are dominated by the 
Anomaly Mediated Supersymmetry Breaking (AMSB) 
contribution \cite{AMSB1,AMSB2}, a situation that 
is naturally realized if the hidden sector does not have a singlet.  In AMSB, 
the masses of the gauginos are suppressed by a loop factor relative to the 
gravitino mass, $m_{3/2}$.  Then the dark matter calculation above sets 
$m_{3/2} \sim 1000 \TeV$.  If the scalar masses are 
unsuppressed relative to this scale, they will heavy enough to satisfy constraints from 
dimension-5 proton decay, flavor changing neutral currents and 
electric dipole moments, making this a theoretically attractive, but experimentally difficult
implementation of split supersymmetry\footnote{In Ref.~\cite{Wells}, a similar 
construction was utilized, but it was assumed that the dark matter was 
generated non-thermally.  This allowed the $\wino$'s to be lighter, with 
possible interesting collider phenomenology.  However, this approach 
is at odds with our philosophy that the thermal abundance of the 
Dark Matter is what puts the fermions at the weak scale. Some discussion 
of the phenomenology of AMSB with unsuppressed scalar masses
was also contained in the earlier \cite{AMSB2}.}.

\section{Direct Detection}
In the MSSM, it is somewhat complicated to discuss the prospects for 
direct detection. There are several intermediate particles that can contribute
to the spin-independent cross-section: the light 
Higgs boson, the heavy Higgs boson, and squarks.  The frustrating possibility 
of a cancellation between various intermediate states exists; in 
principle, the cross-section for direct detection can go nearly to zero\cite{Mandic}.  
In split supersymmetry, there is a single dominant diagram 
for direct detection:  
the exchange of the Higgs boson, utilizing the 
$\bino-\higgsino-h$  or $\wino-\higgsino-h$ 
vertices. These couplings provide a coherent contribution to the 
spin-independent scattering amplitude off nuclei.   

We display the prospects for direct detection of bino-like Dark Matter 
in split supersymmetry in Fig.~2.  We show four sets of points corresponding 
to the models scanned in Fig.~1.

\begin{figure}
\begin{center}
\epsfig{file=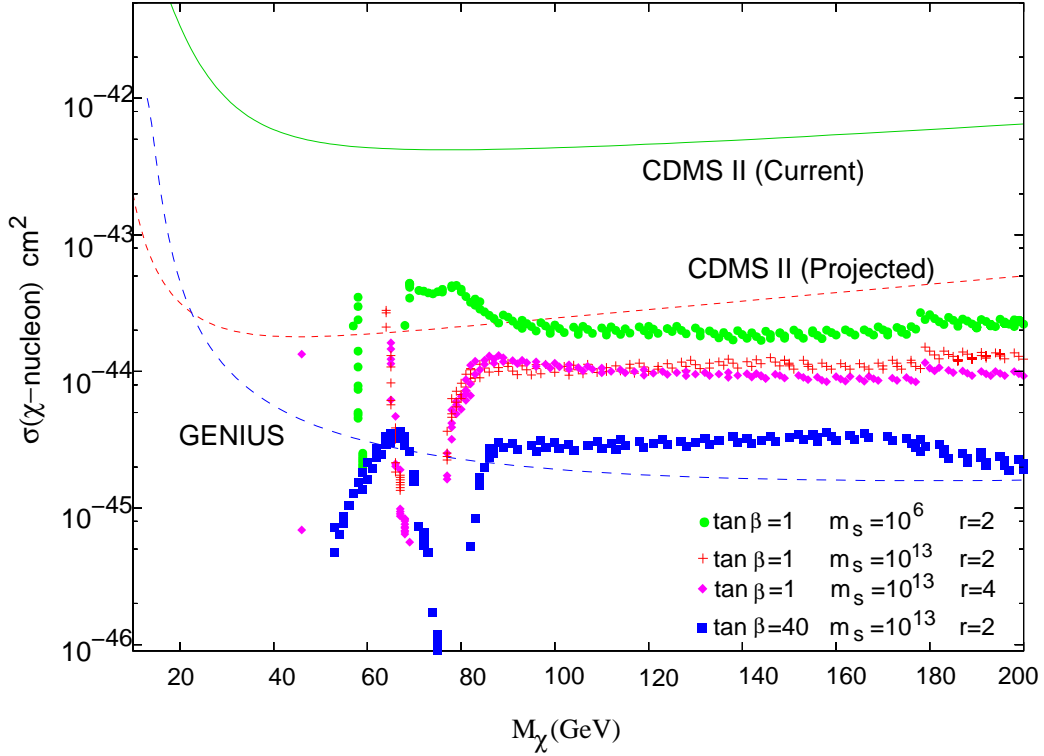,width=\columnwidth}
\label{fig:dd}
\caption{The spin-independent LSP-nucleon scattering cross-section
for point agreeing with the WMAP relic density constraint.    
Curves are plotted for various values of $m_{S}$ (the scale
of the scalar masses), $\tan \beta$ (set at $m_{S}$), and 
$r\equiv M_{2}/M_{1}$.   
The flat regions for $m_{\chi} \gsim 80$ GeV correspond to 
mixed bino-higgsino dark matter.  The sharp decrease at lower masses is due 
to resonant annihilation through the Higgs boson and $Z$-boson poles.  
Also shown are the current bound from the results of 
the Cryogenic Dark Matter Search (CDMS) at Soudan \cite{CDMS}, the
projected bounds from this experiment \cite{CDMSProj}, and projected
bounds from a representative next generation dark matter experiment, 
GENIUS \cite{Genius}}.
\end{center}
\end{figure}

First we discuss some general features.  
In the bino-like dark matter case, there is always a heavy higgsino 
component, particularly in the diagonal stripe of Fig.~1. 
There, the $\bino-\higgsino$ mixing selects a 
fixed value for the $\chi-\chi-h$ vertex.  The diagonals in 
Fig.~1 correspond to the flat horizontal bands 
seen in Fig.~2 for each model above roughly $m_{\chi} \gsim  80 \GeV$.  

Changing $m_{S}$ from $10^{13} \GeV$
to $10^{6} \GeV$ causes an increase in the scattering 
cross section.  This may 
largely be explained by the change in the Higgs boson mass, which 
increases as $m_{S}$ increases:
\begin{equation}
\sigma_{\chi N} \propto \frac{1}{m_{h}^{4}}. 
\end{equation}
For $m_{S} =10^{6}$ GeV, the Higgs boson 
mass is roughly 140 GeV, while for $m_{S} =10^{13}$ GeV, the Higgs boson mass
is roughly 160 GeV.  For all cases considered here the Higgs boson is 
heavier than in the MSSM; the result is a suppression in the detection rate.
Yet the change in the Higgs boson mass does not 
explain the difference in cross section completely.  Even after correcting 
for this factor, the direct detection rates for $m_{S}=10^{6} \GeV$ 
and $m_{S}=10^{13} \GeV$ 
still can differ by order 10\%.  This is because the 
$\kappa'$ parameters, important for detection, run differently from the 
two $m_{S}$ scales.

In Fig.~2,  it can be seen that the rate for $\tan \beta=$40 is suppressed 
relative to $\tan \beta=1$ case.  In the decoupling 
limit ($\cos \alpha= \sin \beta$, $\sin \alpha=-\cos\beta$ ), the 
amplitude for scattering via the light Higgs 
boson in the MSSM is given by \cite{JKG} 
\begin{equation}
A_{h} \propto (g_{2} N_{12} - g_{Y} N_{11})(N_{14} \sin \beta - N_{13} 
\cos \beta).
\end{equation}
In split supersymmetry, the gauge couplings (and $\beta$'s) are 
replaced at low energies by $\kappa$ parameters. 
However, since the $\kappa$'s do not deviate drastically from the 
supersymmetric values, traditional MSSM formulas are still useful to gain 
intuition.   It is instructive to examine 
the higgsino like limit, where \cite{WellsDD}
\begin{equation}
\sigma_{\chi-nucleon} \propto \big( \frac{1 + \sin 2 \beta}{\mu -M_{1}}
\big)^{2}
\end{equation}
In this limit, we can see that the cross section can be enhanced by roughly a factor of  four relative to the large $\tan \beta$ limit 
at $\tan \beta \approx 1$.  This is in rough agreement with Fig.~2; there is 
some deviation due to the mixed $\bino-\higgsino$ nature of the LSP.  We 
expect $\tan \beta=1$ to give scattering cross sections near maximal for 
this model.

The precipitous drop in the cross section below 80 GeV is due to the presence
of the Higgs resonance region.  In the Higgs resonance region, the LSP
can be very pure bino, which leads to a small LSP coupling to the 
Higgs boson.  While the $s$-channel resonance compensates for this
in the early universe, the resonance effect is not available in direct 
detector experiments.  A second drop is visible for the lowest mass 
points on the the $\tan \beta =40$,  $m_{S}=10^{13} \GeV$, $r=2$
 and  $\tan \beta =1$,  $m_{S}=10^{13} \GeV$, $r=4$ curves.  These are points 
where the $Z$-pole is important for determining the relic abundance.

The glitch in the curves at a LSP mass of $m_{\chi} \sim 175$ GeV is 
physical. As the top threshold opens up, the higgsino fraction, $h_{f}$, 
necessary to maintain the measured relic abundance changes. This, in turn, 
affects the direct detection rate.

Note that for the case where $r=2$, $\tan \beta =1$, and 
$m_{S} =10^{13} \GeV$, there is a region near $m_{\chi} \sim 80$ GeV where 
the cross section increases somewhat.  This region is due to the presence of
Dark Matter that has a non-negligible wino fraction.  In this case, 
the $\kappa$ couplings, rather than the $\kappa'$ couplings can enter.   
Unsurprisingly a similar region is not visible for the $r=4,\tan \beta =1$,
$m_{S} =10^{13} \GeV$.  In this case, the wino fraction of the LSP 
is suppressed.

Cross-sections can reach a few $\times 10^{-44} \text{ cm}^{2}$, 
a level that will be covered soon by direct search experiments such as the 
Cryogenic Dark Matter Search (CDMS) experiment in the Soudan mine 
(see Fig.~2).   A planned upgrade (CDMS III) should reach a 
roughly a factor of three below the CDMS II projection.  Future planned 
experiments, such as GENIUS\cite{Genius} could reach the 
$10^{-45} \text{cm}^2$ level. This would allow coverage of 
much of the horizontal region in Fig.~2. 
Relocating a CDMS-like experiment 
to a site with even lower neutron background than the Soudan mine 
could conceivably reach the $10^{-46} \text{cm}^2$ level\cite{BlasC}. 
In this case, even parts of the resonance regions would be accessible.

For the case of very pure higgsino and very pure wino dark matter, the 
direct detection cross section is strongly suppressed.  Since the dominant 
contribution to the spin-independent cross-section comes from the 
$\higgsino-\bino-h$ and $\higgsino-\wino-h$ vertices, non-mixed dark matter 
is difficult for detection.  Unfortunately, scattering cross-sections can be 
astonishingly low: $\sigma_{\chi N} \lsim 10^{-50} \text{ cm}^{2}$.

This study also has interesting implcations 
for dark matter detection in the traditional MSSM with low
energy supersymmetry.  Consider the generic case where the dark matter 
of the MSSM is mixed $\higgsino-\bino$, and resonances and coannihilations 
are unimportant for determining the relic abundance.  
Becuase the Higgs boson mass is heavier
in split symmetry than in the MSSM, in the absence of accidental 
cancellations between detection diagrams, the horizontal band 
with $m_{S}=10^{13}, \tan \beta=40$ of Fig.~2 
actually represents a very conservative {\it lower} bound on the detection 
cross section for MSSM dark matter.  Thus, the cross section 
$10^{-45} \text{ cm}^{2}$ represents an exciting target for 
direct detection both for split supersymmetry and the usual MSSM.

\section{Conclusions}
We have explored dark matter in split supersymmetry.  It is exciting 
that there is a large region of bino-like dark matter where the LSP is 
light.  This could allow for the discovery of charginos and 
neutralinos at the LHC.  There do also exist 
regions that are troublesome for colliders and direct detection: both 
pure higgsino and pure wino dark matter would be quite heavy $(\gsim 1 \TeV)$, 
and would have a very small scattering cross-section off nuclei.
  
It would be interesting to explore the indirect detection of the 
dark matter in this model, especially in this troublesome region 
where direct detection becomes more difficult.  In the pure wino case, 
it might be possible to observe 
the annihilation  of the dark matter to $\gamma \gamma$ at a future 
experiment such as GLAST (Gamma Ray Large Area Space Telescope) \cite{Ullio}. 

Away from the Higgs resonance, the model can have 
mixed bino--higgsino dark matter which should be detectable at 
future direct detection experiments.   It is especially encouraging that a 
region exists where the dark matter could be 
observed at both the LHC and direct detection experiments.  This would allow
for a conclusive demonstration that the particle seen at accelerator
is in fact responsible for the Dark Matter.

\section*{Acknowledgments}
Thanks to A.~Birkedal, B.~Cabrera, P.~Graham, G.~Kribs, M.~Peskin, and 
S.~Thomas for useful discussions, and to 
E.~A.~Baltz and J.~Edjs\"{o} for their assistance with 
the DarkSUSY package.  Special thanks to N.~Arkani-Hamed and 
S.~Dimopoulos for several conversations and collaboration in the 
early stages of this work.

\section*{Note Added}
While this work was being completed, \cite{Giudice} appeared, which has some 
overlap.  It does not discuss the Higgs resonance region.

\end{document}